\documentclass[aps,nofootinbib,prd,eqsecnum,showpacs,showkeys,preprintnumbers]{revtex4-2}
\usepackage[caption=false]{subfig}
\usepackage{graphicx}
\usepackage{amsmath}
\usepackage{amsfonts}
\usepackage{amsmath}
\usepackage{amssymb}
\usepackage{color}
\usepackage{bm}
\usepackage{mathrsfs}
\usepackage{epstopdf}
\usepackage{url}
\usepackage{footnote}
\usepackage{textcomp}
\usepackage{ulem}
\usepackage{esint}
\usepackage[unicode=true, pdfusetitle,
bookmarks=true,bookmarksnumbered=false,bookmarksopen=false,
 breaklinks=false,pdfborder={0 0 1},backref=false,colorlinks=false]{hyperref}
\usepackage{multirow}
\usepackage{pifont}
\begin{document}

\title{Constraints on Power Law and Exponential models in $f(Q)$ Gravity}

\author{Dalale Mhamdi\textsuperscript{1}}
\email{dalale.mhamdi@ump.ac.ma}
\author{Farida Bargach\textsuperscript{1}}
\email{farida.bargach@gmail.com}
\author{Safae Dahmani\textsuperscript{1}}
\email{dahmani.safae.1026@gmail.com}
\author{Amine Bouali\textsuperscript{1,2}}
\email{a1.bouali@ump.ac.ma}
\author{Taoufik Ouali\textsuperscript{1}}
\email{t.ouali@ump.ac.ma}
\date{\today}
\affiliation{\textsuperscript{1}Laboratory of Physics of Matter and Radiations, Mohammed I University, BP 717, Oujda, Morocco,\\  \textsuperscript{2}Higher School of Education and Training, Mohammed I University, BP 717, Oujda, Morocco}

	\begin{abstract}
In this paper, we observationally test the \( f(Q) \) gravity model at both background and perturbation levels using Pantheon$^+$, Hubble measurements, and Redshift Space Distortion Data. We obtain the best-fit parameters by solving numerically the modified Friedmann equations for two distinct cosmological models of \( f(Q) \) gravity namely the Power law and Exponential models. This involves performing a Markov Chain Monte Carlo analysis for these specific forms of \( f(Q) \). To evaluate the statistical significance of the \( f(Q) \) gravity models, we use the Bayesian and corrected Akaike Information Criteria. Our results indicate that the Exponential model in \( f(Q) \) gravity is statistically favored over both the Power-law model and the \( \Lambda \)CDM model.\\
\end{abstract}

	\keywords{$f(Q)$ gravity, MCMC, numerical methods }
	\maketitle
          \tableofcontents
\newpage

\section{Introduction}
The understanding of the Universe's Expansion has been profoundly influenced by astronomical observations, such as those of type Ia supernovae \cite{Riess1998, Perlmutter1999,Riess2004}, cosmic microwave background radiation \cite{Huang2006,Komatsu2011}, and large-scale structures \cite{Koivisto2006,Daniel2008}. These observations reveal that the Universe transitioned from a decelerating phase in its early history to an accelerating phase in more recent times. Identifying the cause of this late-time cosmic acceleration is one of the most significant challenges in modern cosmology. The accelerated Expansion of the Universe finds a compelling explanation in the presence of dark energy, characterized by its negative pressure \cite{Steinhardt1999,Ratra1988}. Among the range of dark energy models \cite{ouali1,safae1,dalale1,amine1,amine2,yashi}, $\Lambda$CDM has proven particularly effective in elucidating crucial cosmological phenomena, such as the formation of large-scale structures and the behavior of type Ia supernovae. Nonetheless, its integration encounters notable theoretical hurdles, particularly concerning challenges related to fine-tuning \cite{Sahni2000,Peebles2003,Weinberg1989} and the cosmic coincidence problem \cite{Sivanandam2013,Velten2014}.\\
Another approach for addressing the mystery surrounding dark energy involves exploring alternative theories of gravity that diverge from Einstein's general theory of relativity (GR). These alternative frameworks hold the potential to account for the Universe's accelerating Expansion without relying on the existence of dark energy. In GR, curvature is quantified by the Ricci scalar, $R$, using principles of Riemannian geometry. However, in modified theories like $f(R)$ gravity, the Ricci scalar is substituted with more flexible functions of $R$ \cite{Capozziello2007}. Furthermore, alternative theories such as $f(T)$ gravity \cite{DeFelice2010,Starobinsky1980} describe gravitational effects using the concept of torsion $T$. These new alternative  theories, which are more general framework than Riemannian geometry, are based on Weyl geometry  and dubbed the
Teleparallel Equivalent of GR (TEGR) \cite{Maluf2013}. In the framework of TEGR, the Weitzenböck connection is used, which yields a spacetime with zero curvature $R$ and zero non-metricity $Q$, but introduces non-zero torsion $T$. TEGR was used to explain the late-time acceleration of the Universe \cite{BengocheaFerraro2009,Linder2010} and to provide an alternative to inflation \cite{FerraroFiorini2007}. To further alleviate the restriction on non-metricity while concurrently eradicating both curvature and torsion, a pioneering theory known as the Symmetric Teleparallel Equivalent of GR (STEGR) has been developed \cite{NesterYo1999}. In this theory, it is conceivable to develop $f(Q)$ gravity, where the non-metricity $Q$ represents the gravitational interaction \cite{Jimenez2018,Jimenez2020}.  In $f (Q)$ gravity, the action is defined by $S =\int \sqrt{-g}f(Q)d^{4}x$ \cite{Jimenez2020}. Furthermore, the modified theory of $f(Q)$ gravity leads to interesting cosmological phenomenology at the background level \cite{Jimenez2020, Dialektopoulosetal2019,Bajardietal2020,FlathmannHohmann2021,Mandaletal2020,D'Ambrosioetal2020,Kossour,omar}. Moreover, it has been effectively tested against diverse observational data, encompassing both background and perturbation observations. These include the Cosmic Microwave Background (CMB), Supernovae type Ia (SNIa), Baryonic Acoustic Oscillations (BAO), Redshift Space Distortion (RSD), as well as growth data, among others, \cite{dalale2,dalale3,Soudietal2019,Lazkozetal2019,Barrosetal2020,Ayusoetal2021,Anagnostopoulosetal2021}, and this confrontation reveals that $f(Q)$ gravity may challenge the standard $\Lambda$CDM scenario. Finally, $f(Q)$ gravity comfortably passes the Big
Bang Nucleosynthesis (BBN) constraints too \cite{Anagnostopoulosetal2022}.\\
Motivated by the intriguing properties of \( f(Q) \) gravity, we investigate two studied models from the literature: the Power-law and the Exponential models. The Power-law form, as given by \cite{Jimenez2020, Lazkozetal2019, Ayusoetal2021}, is expressed as \( F(Q) = \alpha \left(\frac{Q}{Q_0}\right)^\lambda \). The Exponential model, described by \cite{Anagnostopoulosetal2021}, takes the form \( F(Q) = \alpha Q_0 \left(1 - e^{\left(-\beta \sqrt{\frac{Q}{Q_0}}\right)}\right) \). These models lack analytical solutions for the modified Friedmann equations, necessitating numerical approaches. In this paper, we aim to constrain \( f(Q) \) gravity numerically through Bayesian analysis at both background and perturbation levels. We estimate the model parameters using data from various probes, including redshift space distortions \cite{Turnbulletal2012, Achitouvetal2017, Beutleretal2012, Feixetal2015}, Hubble measurements \cite{Zhangetal2014, Jimenezetal2003, Simonetald2005, Morescoetal2012, Gaztanagaetal2009}, and Pantheon$^+$ datasets \cite{Broutetal2022}. Subsequently, we employ a Markov Chain Monte Carlo (MCMC) analysis \cite{Mehrabietal2015,MCMC1} to extract the best-fit values of the free parameters for both models. With these parameter values constrained, our analysis yields several noteworthy findings.\\
The present paper is organized as follows. Firstly, in Sec. \ref{sec:3}, we introduce the formalism of $f(Q)$ gravity. Then, in Sec. \ref{fq}, we present two models: Power-law and exponential. In Sec. \ref{section.4}, we describe the different observational datasets used to constrain the $f(Q)$ model parameters, discuss the use of the MCMC method, and present the probability function. We present and discuss our results in Sec. \ref{results}. Finally, the obtained results are summarized in Sec. \ref{conclusion}.

 \section{$f(Q)$ cosmology} \label{sec:3}
In this section, we discuss the detailed formulation of the symmetric teleparallelism, specially its extension the modified $f(Q)$ theory. To initiate the discussion, we define the action of $f(Q)$ gravity as presented in \cite{JimenezHeisenbergKoivisto2018},
\begin{equation}
S = \int d^4x \sqrt{-g} \left[ \frac{f(Q)}{16\pi G} + \mathcal{L}_m \right].
\end{equation}
In this model, $\mathcal{L}_m$ represents the matter Lagrangian, and $g$ denotes the determinant of the metric tensor $g_{\mu\nu}$. Furthermore, $f(Q)$ is an arbitrary function of the nonmetricity $Q$. The non-metricity tensor $Q_{\gamma\mu\nu}$, which is the
covariant derivative of the metric tensor with respect to
the Weyl-Cartan connection, is discussed in \cite{Hehletal1976},
\begin{align}
 Q_{\gamma \mu \nu} = \nabla_\gamma g_{\mu \nu}=-\partial_{\gamma}g_{\mu\nu}+g_{\nu\sigma}\Upsilon^\sigma_{\mu\gamma}+g_{\sigma\mu}\Upsilon^{\sigma}_{\nu\gamma}.   
\end{align}
The Christoffel symbol $\Gamma^\gamma_{\mu\nu}$, the contortion tensor $C^\gamma_{\mu\nu}$, and the disformation tensor can be combined
to form the Weyl-Cartan connection. This connection is defined in the following way 
\begin{align}\label{eq1}
 \Upsilon^\gamma_{\ \mu\nu} = \Gamma^\gamma_{\ \mu\nu} + C^\gamma_{\ \mu\nu} + L^\gamma_{\ \mu\nu}.   
\end{align}
The first term in Eq. \eqref{eq1}, the Christoffel symbol, corresponds to the Levi-Civita connection
associated with the metric $g_{\mu\nu}$. It is given by the following expression
 \begin{align} 
     \Gamma^\gamma_{\ \mu\nu} \equiv \frac{1}{2} g^{\gamma\sigma} \left( \partial_\mu g_{\sigma\nu} + \partial_\nu g_{\sigma\mu} - \partial_\sigma g_{\mu\nu} \right).
 \end{align}
The torsion tensor $T^\gamma_{\ \mu\nu}$ is employed in the construction of the contortion tensor $C^\gamma_{\ \mu\nu}$, which is given by
\begin{align}
C^\gamma_{\ \mu\nu} \equiv \frac{1}{2} (T^\gamma_{\ \mu\nu} + T^\gamma_{\ \mu\nu} + T^\gamma_{\ \nu\mu}) = -C^\gamma_{\ \nu\mu}.    
\end{align}
The disformation tensor, which is constructed from the non-metricity, is given by

\begin{align}
    L^\gamma_{\ \mu\nu} =& -\frac{1}{2} g^{\gamma\sigma} \left( \nabla_\nu g_{\mu\sigma} + \nabla_\mu g_{\nu\sigma} - \nabla_\gamma g_{\mu\nu} \right),\\
=& \frac{1}{2} g^{\gamma\sigma} \left( Q^\nu_{\ \mu\sigma} + Q^\mu_{\ \nu\sigma} - Q^\gamma_{\ \mu\nu} \right),\\
= &L^\gamma_{\ \nu\mu}.
\end{align}
Thus, the non-metricity can take the following expression
\begin{align}
    Q = - g^{\mu\nu} \left( L^\alpha_{\ \beta\mu} L^\beta_{\ \nu\alpha} - L^\alpha_{\ \beta\alpha} L^\beta_{\ \mu\nu} \right).
\end{align}
The trace of the non-metricity tensor can be expressed as
\begin{align}
    Q_\alpha = Q_\alpha^{\ \mu}{}_\mu \quad , \quad \tilde{Q}_\alpha = Q^{\mu}_{\alpha}{}_\mu.
\end{align}
It is also useful to introduce the superpotential tensor (the conjugate of non-metricity) defined by
 \begin{align}
    4P^\gamma_{\ \mu\nu} = - Q^\gamma_{\ \mu\nu} + 2 Q_{(\mu}{}^{\gamma}{}_{\nu)} + (Q^\gamma - \tilde{Q}^\gamma) g_{\mu\nu} - \delta^{\gamma}_{(\mu} Q_{\nu}),
\end{align}

 where the trace of the nonmetricity tensor can be obtained as
 \begin{align}
     Q = -Q_{\gamma \mu \nu}P^{\gamma \mu \nu}.
 \end{align}
By varying the action $S$ with respect to the metric tensor $g_{\mu\nu}$, we obtain the field equations of the symmetric teleparallel gravity
\begin{align}
   \frac{2}{\sqrt{-g}} \nabla_{\gamma} (\sqrt{-g}f_Q P^{\gamma}_{}{\mu \nu}) + \frac{1}{2} g_{\mu \nu} f + f_Q(P_{\nu\rho\sigma}Q_\mu^{}{\rho\sigma}-2P_{\rho\sigma\mu} Q^{\rho\sigma}_{}{\nu})= T_{\mu \nu},
\end{align}
where the energy-momentum tensor is given by
\begin{align}
    T_{\mu \nu} = -\frac{2}{\sqrt{-g}} \frac{\delta (\sqrt{-g} \mathcal{L}_{m})}{\delta g^{\mu \nu}}.
\end{align}
By varying the action with respect to the connection, we get
\begin{align}
    \nabla^{\mu} \nabla^{\nu} (\sqrt{-g} f_Q P^{\gamma} _{}{\mu \nu}) = 0.
\end{align}
We assume a homogeneous, isotropic and spatially flat Friedmann-Lemaître-Robertson-Walker (FLRW) spacetime, whose metric is
of the forme
\begin{align}
   \mathrm{d} s^2=-\mathrm{d} t^2+a^2(t)\left[\mathrm{d} r^2+r^2\left(\mathrm{~d} \theta^2+\sin ^2 \theta \mathrm{d} \phi^2\right)\right]
\end{align}
where $a(t)$ represents the scale factor. Moreover, the non-metricity scalar corresponding to the FLRW metric can be expressed as
\begin{align}
    Q=6H^2,
\end{align}
where $H$ represents the Hubble function. The energy-momentum tensor of a perfect fluid is given by
\begin{align}
    T_{\mu \nu} = (p + \rho)u_{\mu}u_{\nu} + p g_{\mu \nu},
\end{align}
where $\rho$ and $p$ represent the energy density and isotropic pressure of the perfect cosmic fluid, respectively, and $u^\mu = (1, 0, 0, 0)$ denotes
the four-velocity components of the cosmic perfect fluid. The modified Friedmann equations that govern the dynamics of the Universe in $f(Q)$ gravity are given by \cite{Harko2018,Lazkoz2019}
\begin{align}\label{friedmann1}
 3H^2 = \frac{1}{2f_Q}\left( -\rho + \frac{f}{2}\right),
 \end{align}
 \begin{align}\label{friedmann2}
 \dot{H} + 3H^2 + \frac{\dot{f_Q}}{f_Q} H = \frac{1}{2f_Q}  \left( p + \frac{f}{2} \right),  
 \end{align}
 where an overdot denotes the derivative with respect
to cosmic time, $t$. Imposing the splitting $f(Q) = Q + F(Q)$, Eqs. \eqref{friedmann1} and \eqref{friedmann2} can be expressed as
\begin{align}
    3H^2 = \rho + \frac{F}{2} - QF_Q,
    \label{fr1}
\end{align}
\begin{align}
    (2QF_{QQ} + F_Q +1)\dot{H}+ \frac{1}{4} (Q + 2QF_Q - F )= -2p,
\end{align}
where $F_Q=\frac{dF}{dQ}$ and $F_{QQ}=\frac{d^2F}{dQ^2}$. In the following section, we will explore two $F(Q)$ models: the Power-law model and the exponential model.

\section{$F(Q)$ models}\label{fq}


\subsection{Power-law $F(Q)$ model}
First, we investigate the Power-law model $F(Q)$, which is represented as \cite{Gadbail,Jimenez2020,Khyllep,Lazkozetal2019,Ayusoetal2021}
\begin{equation}
F(Q)=\alpha\left(\frac{Q}{Q_0}\right)^\lambda,
\label{f1}
\end{equation}
where \(\lambda\) and \(\alpha\) are model parameters. For \(\alpha = 0\), the model reduces to the symmetric teleparallel theory equivalent to GR. When \(\lambda = 0\), the model recovers the \(\Lambda\)CDM model. Using Eqs. (\ref{fr1}) and (\ref{f1}), we obtained

\begin{equation}
E^{2}+\frac{\alpha}{Q_0}(2 \lambda-1)E^{2\lambda}=\Omega_{m} a(t)^{-3},
\label{so1}
\end{equation}
where $\Omega_{m}$ is the present matter density, $E^{2} = \frac{H^{2}}{H_{0}^{2}}$ represents the normalized Hubble parameter where $H_0$ is the current Hubble constant, and $Q_0 = 6H_{0}^2$. Using Eq. (\ref{so1}), the parameter $\alpha$ at the present time ($z=0$) is obtained as

\begin{equation}
\alpha=\frac{\left(\Omega_{m}-1\right)}{2 \lambda-1} Q_0,
\end{equation}
where we have set $a_0=1$. Using the expression of $\alpha$, Eq. \ref{so1} further reduces
\begin{equation}
E^2+\left(\Omega_{m}-1\right)E^{2\lambda}= \Omega_{m} a(t)^{-3}.
\label{Model1}
\end{equation}

\subsection{Exponential $F(Q)$ model}
Next, we consider the exponential $F(Q)$ model, which is of the form \cite{Anagnostopoulosetal2021,Gadbail}

\begin{equation}
F(Q)=\alpha Q_0\left(1-e^{\left(-\beta \sqrt{\frac{Q}{Q_0}}\right)} \right),
 \label{f2}
\end{equation}

where $\alpha$ and $\beta$ are model parameters. For $\beta=0$ the model reduces to the symmetric teleparallel theory equivalent to GR without a cosmological constant. 
Using Eqs (\ref{fr1}) and (\ref{f2}), we obtained
\begin{equation}
E^{2} - \alpha \left[1 - (1 + \beta E) e^{-\beta E}\right] = \Omega_{m} a(t)^{-3},
\label{s2}
\end{equation}
where $\Omega_{m}$ is the present matter density. Using the above equation, the value of model parameter $\alpha$ at the present time is obtained as
\begin{equation}
\alpha=\frac{1-\Omega_{m}}{1-(1+\beta) e^{-\beta}}.
\end{equation}
Using the expression of $\alpha$, Eq. \ref{s2} further reduces to

\begin{equation}
    E^{2} -(1-\Omega_{m}) \frac{1 - (1 + \beta E) e^{-\beta E} }{1-(1+\beta) e^{-\beta}} = \Omega_{m} a(t)^{-3}.
    \label{Model2}
\end{equation}

\section{Cosmological constraints}\label{section.4}

In this Section, we test the observational viability of the two previously introduced cosmological models of $f(Q)$ gravity. To this aim, a detailed statistical analysis comparing the theoretical predictions of the $f(Q)$ gravity models with the cosmological observations is performed. More precisely, the parameters of the two models, namely, ($\Omega_m$, $\lambda$, $h$) for the Power-law model and ($\Omega_m$, $\beta$, $h$) for the exponential model, are constrained by cosmological observations, using the Markov Chain Monte  Carlo approach \cite{Mehrabietal2015,MCMC1}. It should be noted that the parameter $h$ is related to the current Hubble rate as $h = H_{0}/100$ km s$^{-1}$ Mpc$^{-1}$. The observational data sets used to analyse the cosmological models in $f(Q)$ gravity are Pantheon$^+$ \cite{Broutetal2022}, Hubble measurements \cite{Zhangetal2014, Jimenezetal2003, Simonetald2005, Morescoetal2012, Gaztanagaetal2009}, and redshift space distortion \cite{Turnbulletal2012, Achitouvetal2017, Beutleretal2012, Feixetal2015}.\\

\subsection{Methodology}
Nowadays, cosmological theories are evaluated through various observational data gathered from a range of cosmological probes, including Type Ia supernovae, baryon acoustic oscillations (BAO), the cosmic microwave background (CMB), redshift space distortions (RSD), and Hubble measurements. Confronting theoretical models with observations is the only way to distinguish between observationally supported, and unsupported models. To efficiently compare theoretical models with the observational data, a detailed statistical analysis is required, involving a multilevel approach. As cosmology is Bayesian by its very nature, parameters estimation processes are, in general, performed in the Bayesian inference framework

\begin{equation}
\mathcal{P}(\theta \mid D) \propto \mathcal{L}(D \mid \theta),
\end{equation}

where $\mathcal{P}(\theta \mid D)$ is the posterior distribution, and $\mathcal{L}(D \mid H)$ is the likelihood function. In the case of Gaussian errors, the chi-square function, $\chi^{2}$, and the likelihood function, $ \mathcal{L}$, are related according to the relation
\begin{equation}
    \chi^2(\theta) = -2 \ln \mathcal{L}(\theta).
\end{equation}

With the aim of estimating the parameters of a given theoretical model, one has to look for the free parameters vector, $\theta$, which minimizes the chi-square function, or, equivalently, maximizes the likelihood function. To constrain the cosmological parameters of the $f(Q)$ gravity models, we minimize the $\chi^{2}_{tot}$ function that is expressed as  

\begin{align}
    \chi^2_{tot} = \chi^2_{\text{Pantheon}^{+}}  + \chi^2_{H(z)}+ \chi^2_{RSD}.
\end{align}

To evaluate and select the most appropriate model for explaining observational data, a comparison criterion is usually adopted. In  this analysis, we use the corrected Akaike Information Criterion (AIC$_{c}$) \cite{Rosa} and the Bayesian Information Criterion (BIC) \cite{Vrieze}, expressed respectively as
\begin{equation}
    AIC_c = \chi^2_{\text{min}} + 2K_f + \frac{2K_f(K_f + 1)}{N_t - K_f - 1},
\end{equation}
and 
\begin{equation}
\mathrm{BIC}=\chi_{\min }^2+\mathcal{K}_{\mathrm{f}} \ln \left(\mathcal{N}_{\mathrm{t}}\right)
\end{equation}
where $\mathcal{K}_{\mathrm{f}}$ and $\mathcal{N}_{\mathrm{t}}$ represent the number of the free parameters, and the total number of the data points employed, respectively. The model with the lowest $AIC_c$ and $BIC$ is the most supported by observations, and it is chosen to be the reference model. In the remainder of this work, we use $\Lambda$CDM as a reference model. To measure how closely models resemble the reference one, we compute the quantities
\begin{equation}
   \Delta AIC_{c} = AIC_{c, \text{model}} - AIC_{c, \text{ref}} ,
\end{equation}
and
\begin{equation}
    \Delta BIC = BIC_{\text{model}} - BIC_{\text{ref}} .
\end{equation}
   For \(0 < |\Delta \text{AIC$_c$}| < 2\), the model has the same level of support from the dataset as the reference model. For \(2 < |\Delta \text{AIC$_c$}| < 4\), the model with the larger AIC$_{c}$ value is less favored by the data. For \(4 < |\Delta \text{AIC$_c$}| < 6\), there is positive evidence against the model with the larger AIC$_{c}$ value. For \(6 < |\Delta \text{AIC$_c$}| < 10\), there is strong evidence against the model with the larger AIC value. For \(10 < |\Delta \text{AIC$_c$}|\), the model with the larger AIC$_{c}$ value is not supported by the data. The same interpretation applies to BIC.

\subsection{Background data}
\begin{itemize}
    \item \textbf{Pantheon$^+$ dataset:}
    Recently, a refreshed version of the Pantheon dataset, referred to as Pantheon$^+$, has been introduced \cite{Broutetal2022}. These datasets, comprising 1701 data points, are obtained from 1550 type Ia supernovae spanning a redshift range of $0.001 \leq z \leq 2.3$. The $\chi^2$ function for the supernovae datasets is presented as follows
\begin{align}
   \chi_{\text{Pantheon$^+$}}^2 = \vec{F}^T \cdot  C^{-1}_{\text{Pantheon$^+$}} \cdot \vec{F}
   \label{chi}
\end{align}
where, $\vec{F}$ represents the vector of the difference between the observed apparent magnitudes $m_{\text{Bi}}$ and the expected magnitudes given by the cosmological model. The term $\mathbf{C}_{\text{Pantheon$^+$}}$ denotes the covariance matrix provided with the Pantheon$^+$ data, which includes both statistical and systematic uncertainties.
The distance modulus is a measure of the distance
to an object, defined as follows 

\begin{align}
   \mu_{\text{model}}(z_i) = 5 \log_{10} D_L(z_i) + 25.
\end{align}
where $D_L$ represents the luminosity distance, expressed as
\begin{align}
    D_L(z_i) = (1 + z) \int_0^{z} \frac{c}{H(z')} dz',
\end{align}
with $c$  representing the speed of light. The Pantheon$^+$ dataset diverges from its predecessor by resolving the interdependence between the absolute magnitude, $M$, and the Hubble constant $H_0$. This is accomplished by redefining the vector $\vec{F}$ in terms of the distance modulus of SNIa in the Cepheid hosts. These moduli are independently measured using Cepheid calibrators provided by SH0ES \cite{SH0ES}, facilitating a separate determination of $M$. Consequently, the adjusted vector $\vec{F}'$ is formulated as
\begin{align}
  \vec{F_i}   = \begin{cases} 
m_{\text{Bi}} - M - \mu_{i}^\text{Ceph} & i \in \text{Cepheid hosts }\\ m_{\text{Bi}} - M - \mu_\text{model}(z_i)& \text{otherwise}
\end{cases}
\end{align}
where $\mu_{i}^\text{Ceph}$ represents the distance modulus associated with the Cepheid host of the $i$th SNIa, independently measured by Cepheid calibrators. $M$ represents the absolute magnitude of SNIa, and $\mu_{\text{model}}$ is the corresponding distance modulus predicted by the assumed cosmological model. Thus, Eq. \eqref{chi} can be re-expressed as follows
\begin{align}
    \chi_{\text{SN}}^2 = \vec{F}^T \cdot \mathbf{C}_{\text{Pantheon$^+$}}^{-1} \cdot \vec{F}.
\end{align}

\end{itemize}
\begin{itemize}
    \item\textbf{Hubble dataset:} Expressed as $H(z) = -\frac{dz}{dt}(1 + z)^{-1}$, the Hubble parameter allows for an independent determination of its value from observational data. Obtaining $dz$ from a spectroscopic survey and measuring $dt$ facilitates the calculation of the Hubble parameter without reliance on a specific cosmological model. The chi-square function for the Hubble measurements is derived according to the following definition
\begin{equation}
\chi_{\text{H(z)}}^2  = \sum_{i=1}^{57} \left[\frac{H_{obs}(z_i)-H_{th}(z_i)}{\sigma(z_i)}\right]^2,
\end{equation}
where $H_{obs}$ and $H_{th}$ represent the observed and the theoretical values of the Hubble parameter, respectively. Furthermore, $\sigma(z_i)$ corresponds to the error on the observed value  of the Hubble parameter $H(z)$. We use the data reported in \cite{AmineHarko}, which includes a total of  $\mathcal{N}_{\mathrm{t}}  =57$ measurements.
 
\end{itemize}
\begin{table}[t!]
    \centering
    \begin{tabular}{lcccc}
        \toprule
        Models & Parameters & Priors & Pantheon$^+$+H(z) & Pantheon$^+$+H(z)+RSD \\
        \hline
        \multirow{3}{*}{$\Lambda$CDM} & $\Omega_m$ &[0,1]& 0.2729 $\pm$ 0.0107  & 0.272 $\pm$ 0.0105  \\
                                      & $h$ &[0.4,2]& 0.7023 $\pm$ 0.00638 & 0.702 $\pm$0.006 \\
                                      & $\sigma_8$ &[0,2]& $-$ & 0.8094 $\pm$0.0242 \\
                                      & M & [$-$20,$-$19] & $-$19.362 $\pm$ 0.017 & $-$19.362 $\pm$ 0.018\\
                                      
        \hline
        \multirow{4}{*}{Power-law}    & $\Omega_m$ &[0,1] & 0.240$^{+0.022}_{-0.019}$ & 0.256 $\pm$ 0.014 \\
                                      & $h$ &[0.4,2] &0.697$\pm$ 0.007& 0.6986$\pm$0.0068\\
                                      & $\sigma_8$ &[0,2]& $-$ & 0.866$^{+0.039}_{-0.051}$ \\
                                      & $\lambda$ & [0,2]& 0.27 $\pm$ 0.11  & 0.172 $\pm$0.079 \\
                                      & M &[$-$20,$-$19]& -19.370$\pm$ 0.018  & $-$19.269$\pm$0.018\\
        \hline
        \multirow{4}{*}{Exponential}  & $\Omega_m$ &[0,1]& 0.258 $\pm$ 0.013 & 0.259 $\pm$ 0.013 \\
                                      & $h$ &[0.4,2]& 0.6978 $\pm$ 0.0069& 0.6974 $\pm$ 0.0069 \\
                                      & $\sigma_8$  &[0,2]& $-$ & 0.834 $\pm$ 0.029 \\
                                      & $\beta$  &[0,5]& 3.49 $^{+0.62}_{-0.77}$ & 3.53 $^{+0.62}_{-0.77}$ \\
                                      & M & [$-$20,$-$19]& -19.370$\pm$ 0.018  & $-$19.269$\pm$0.018\\
        \hline
    \end{tabular}
    \caption{Mean value \(\pm\) 1\(\sigma\) of parameters from different models using Pantheon$^+$+H(z) and Pantheon$^+$+H(z)+RSD datasets.
}
    \label{bf}
\end{table}

\subsection{Perturbation data}
\begin{itemize}
    \item \textbf{Redshift Space Distortion dataset:} One of the data  we used in this work is the growth rate data compilation, specifically the value of the growth rate \( f(z) \) multiplied by the amplitude of the matter power spectrum on the scale of \( 8h^{-1} \) Mpc, \( \sigma_{8}(z) \), usually written as \( f\sigma_{8}(z) \). We consider a total of \(  \mathcal{N}_{\mathrm{t}}  = 20 \) data points from \cite{dalale2} for different redshifts, \( 0.02 \leq z \leq 1.944 \), with the \(\chi^2_{RSD}\) function defined as
    

\begin{align}\label{sigma}
    \chi^2_{RSD} =\sum_{i=1}^{20}\left(\frac{f\sigma_{8,ob}(z_i) - f\sigma_{8,th}(z_i)}{\sigma(z_i)}\right)^2.
\end{align}

In Eq. \eqref{sigma}, $\sigma(z_i)$ represents the standard error associated with the observed value of $f\sigma_{8}$. The terms $f\sigma_{8,ob}$, and $f\sigma_{8,th}$ correspond to the observed and theoretical values of $f\sigma_8$, respectively. Moreover, the growth rate is given by \cite{nesseris2017}
\begin{align}
     f\sigma_{8,th}(z_i) = \sigma_8 \frac{\delta'(z_i)}{\delta(z=0)},
     \label{Pert}
\end{align}
where a prime denotes the derivative with respect to $x=\ln(a)$.
The quantity $\sigma_8$ stands for the amplitude of the matter power spectrum at the present time, i.e.
$z = 0$. To obtain the theoretical prediction from Eq. (\ref{Pert}), it is necessary to determine the values $\delta'(z)$ and $\delta(0)$ by numerically solving the equation of the matter overdensity at the quasi-static limit which represents as follow

\begin{align}\label{delta}
   \ddot{\delta}_m + 2H \dot{\delta}_m - 4\pi G_{\text{eff}} \rho_m \delta_m = 0,
\end{align}

where the dot denotes the differentiation with respect to the cosmic time and $G_{\text{eff}}$ represents the effective gravitational. In order to simplify our calculations, we rewrite Eq. $\ref{delta}$ in terms of $x$, as follows
\begin{equation}
    \delta''_m + \left( \frac{H(x)'}{H(x)} + 2 \right) \delta'_m - \frac{\rho_m}{2 (1+F_Q) H(x)^2} \delta_m = 0.
\end{equation}

\end{itemize}

\begin{table}[t!]
 \begin{tabular}{c|c|c|c|c|c}
        \hline
        Models & $\chi_{min}^{2}$ & AIC & $\Delta$AIC  & BIC &  $\Delta$BIC \\
        \hline
        $\Lambda$CDM & 1604.12 & 1612.14 & 0  & 1634.05 & 0 \\
         \hline
        Power-law & 1601.38 & 1611.41 & $-$0.7287  & 1638.8 & 4.74\\
        \hline
        Exponential & 1599.86 & 1609.89 & $-$2.249 & 1637.28 & 3.22\\
        \hline
    \end{tabular}
    \caption{Comparison of cosmological models using $AIC_c$, and BIC. The table presents these values along with their differences ($\Delta$AIC and $\Delta$BIC) relative to the $\Lambda$CDM model, based on the combined Pantheon$^+$, H(z), and RSD datasets.}

    \label{tab:model_comparison}
\end{table}

\section{Results and discussions}\label{results}

\begin{figure}[t!]
    \centering    \includegraphics[width=0.5\textwidth]{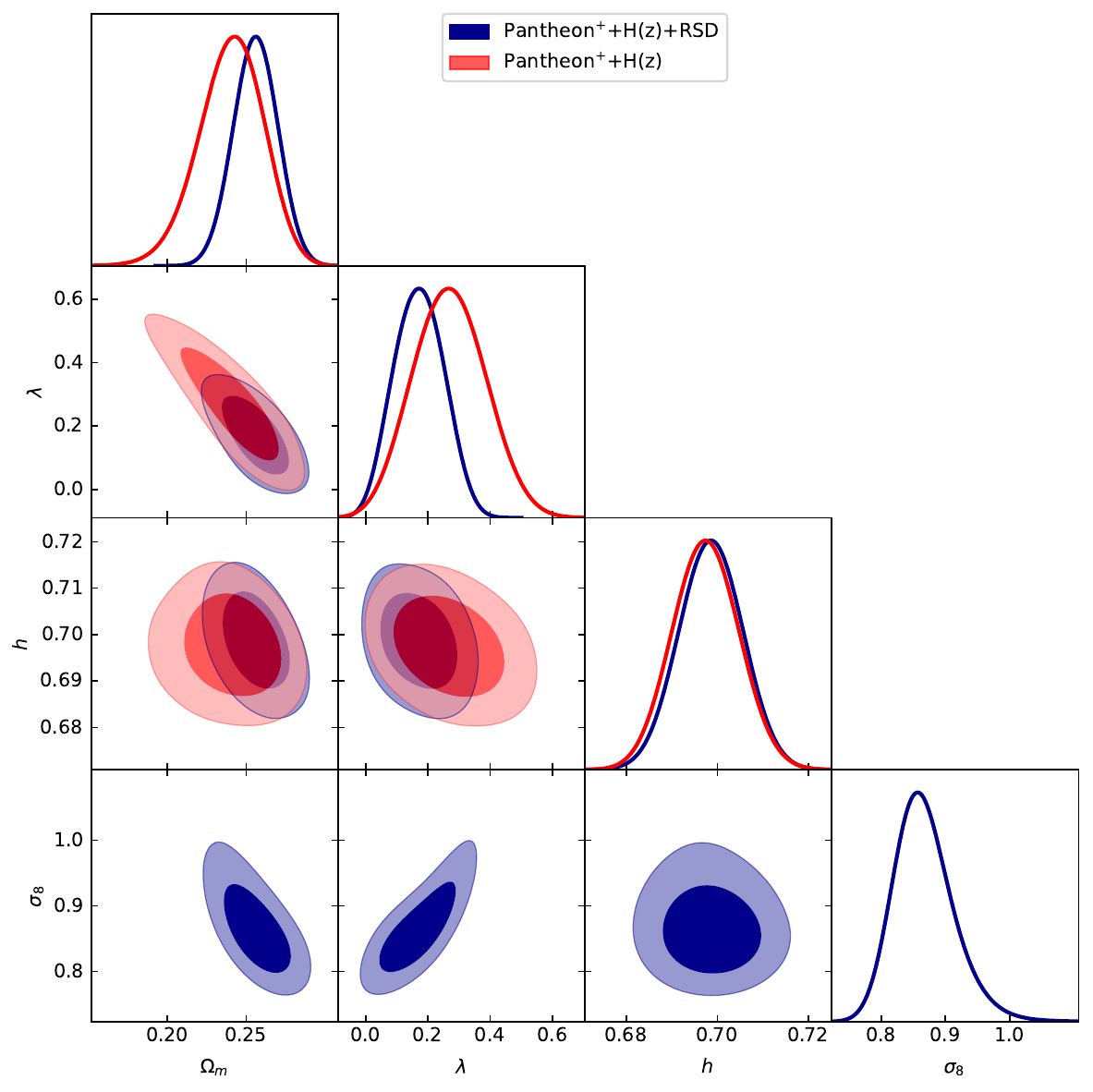}
    \caption{The 1$\sigma$ and 2$\sigma$ confidence contours and the posterior distributions obtained for the $f(Q)$ Power-law model.}
    \label{cc2}
\end{figure}

\begin{figure}[t!]
    \centering
    \includegraphics[width=0.5\textwidth]{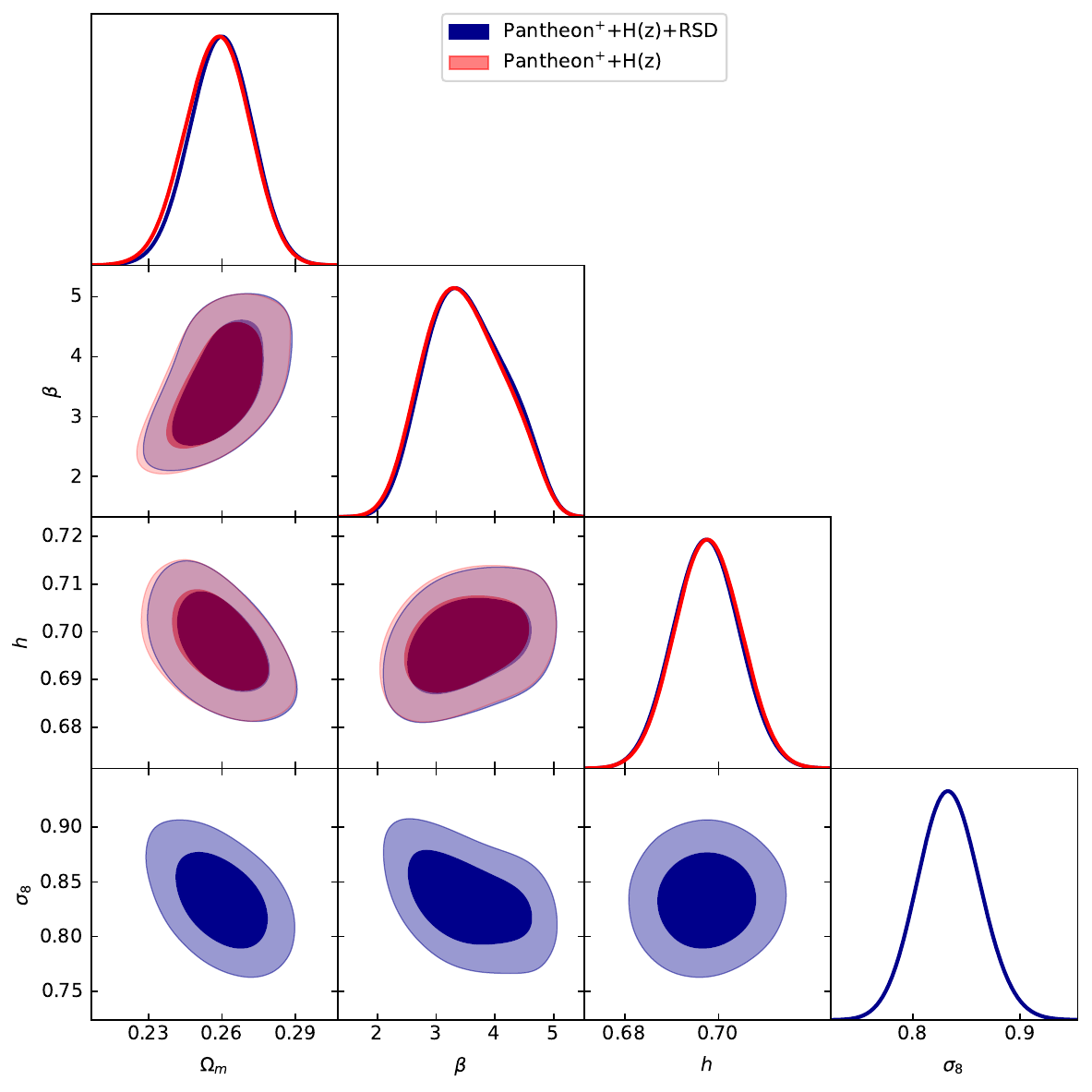}
    \caption{The 1$\sigma$ and 2$\sigma$ confidence contours and the posterior distributions obtained for the $f(Q)$ exponential model.}
    \label{cc3}
\end{figure}

In this section, we present the findings resulting from our analysis of two cosmological models within \( f(Q) \) gravity theory. We test two specific \( f(Q) \) models: the Power-law and exponential models, described by the Friedmann equations \eqref{Model1} and \eqref{Model2}, respectively. As these equations cannot be solved exactly using analytical methods, numerical solutions are required for their computation. We compute the numerical solutions for both models and constrain their free parameters using  Pantheon$^+$, H(z), and RSD. The priors for cosmological parameters. \\

Table \ref{bf} presents the MCMC results for the $\Lambda$CDM, Power-law, and Exponential models. The table includes the prior distributions of the free parameters used in the MCMC analysis, as well as their mean values and associated errors at 1$\sigma$  (68\% confidence level) using two combinations of datasets: Pantheon$^+$+H(z) and Pantheon$^+$ +H(z)+RSD. The free parameter vector for the \(\Lambda\)CDM model consists of $ \{\Omega_m, h, \sigma_8, M\}$. For the Power-law model, the vector includes $\{\Omega_m, h, \lambda, \sigma_8, M\}$, and for the exponential model, it includes $\{\Omega_m, h, \beta, \sigma_8, M\}$.  \\

Constraining the \(\Lambda\)CDM and \( f(Q) \) models with Pantheon$^+$+H(z)+RSD datasets, we find \(\Omega_m = 0.272 \pm 0.0105\) for the \(\Lambda\)CDM model, which is slightly higher than both \( f(Q) \) models: \(\Omega_m = 0.256 \pm 0.014\) and \(\Omega_m = 0.259 \pm 0.013\), for the Power-law model and the Exponential model, respectively. The parameter \( h \) is determined to be \( h = 0.703 \pm 0.006 \, \text{km/s/Mpc} \) for \(\Lambda\)CDM, which is slightly higher compared to $f(Q)$ models where the values are: \( h = 0.698 \pm 0.0068 \, \text{km/s/Mpc} \) and \( h = 0.6987 \pm 0.0069 \, \text{km/s/Mpc} \) for the Power-law and Exponential models, respectively. The amplitude of matter perturbations, \(\sigma_8\), is crucial in describing the clustering of matter on large scales. For \(\Lambda\)CDM, we find \(\sigma_8 = 0.809 \pm 0.024\), which is lower compared to the Power law model (\(\sigma_8 = 0.866^{+0.039}_{-0.051}\)) and the exponential model (\(\sigma_8 = 0.834 \pm 0.029\)).\\


In Table \ref{tab:model_comparison}, we summarize the results of the statistical analysis, highlighting a key finding of this study: the exponential model exhibits statistical superiority over the \(\Lambda\)CDM model, as indicated by \(\Delta \text{AIC}_{c} = -2.249\). This suggests that observations favor the exponential model over the \(\Lambda\)CDM model. However, for the Power-law model, $\Delta \text{AIC}_{c} = -0.728$, which is not sufficient to select it as the best model. Additionally, according to the BIC criterion, the \(\Lambda\)CDM model is statistically favored, primarily due to its simplicity with fewer parameters compared to \( f(Q) \) gravity models.\\

\begin{figure}
    \centering
    \includegraphics[width=0.5\textwidth]{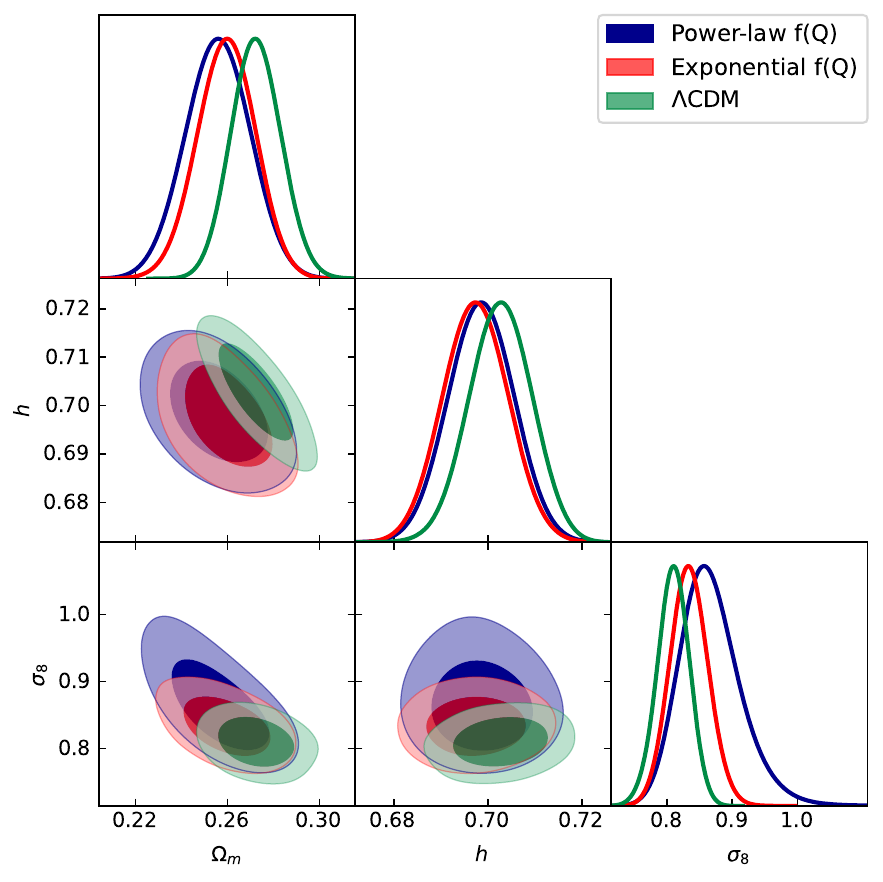}
    \caption{The 1$\sigma$ and 2$\sigma$ confidence contours and the posterior distributions obtained for the $\Lambda$CDM, Power-law, and Exponential models, using Pantheon$^+$+H(z)+RSD}
    \label{CCT}
\end{figure}

\begin{figure}[h!]
    \centering
    \begin{minipage}{0.45\textwidth}
        \centering
        \includegraphics[width=\textwidth]{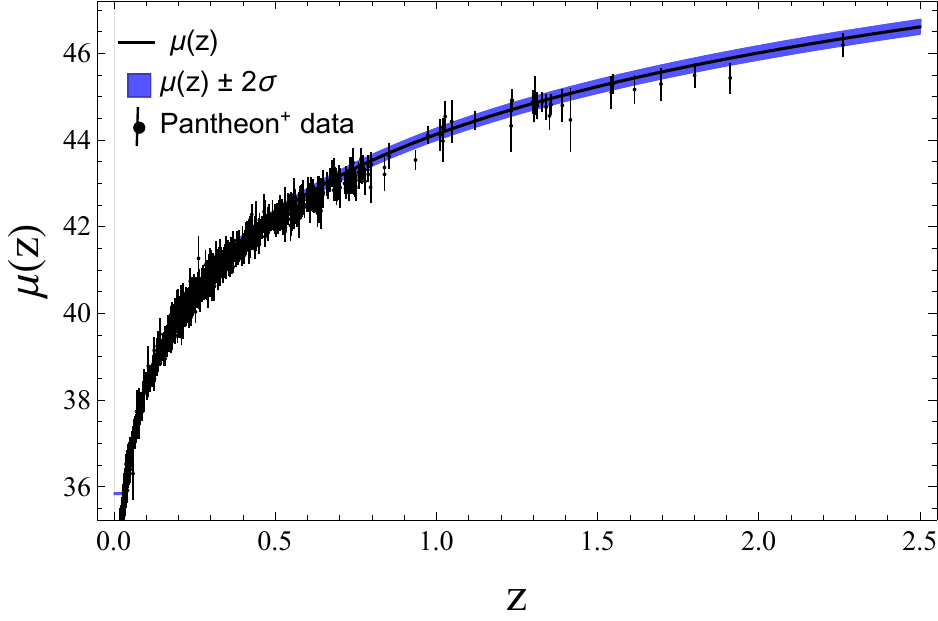}
    \end{minipage}
    \hspace{0.05\textwidth}
    \begin{minipage}{0.45\textwidth}
        \centering
        \includegraphics[width=\textwidth]{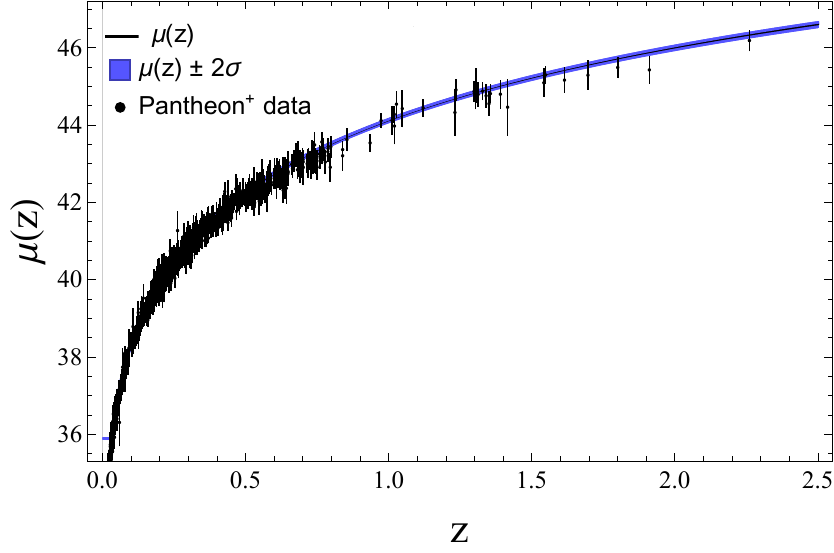}
    \end{minipage}
    \caption{The figures represent the predicted $\mu(z)$ function with 2$\sigma$ error for Power-law (left) and Exponential (right) models.}
    \label{fig:figures1}
\end{figure}

\begin{figure}[h!]
    \centering
    \begin{minipage}{0.45\textwidth}
        \centering
        \includegraphics[width=\textwidth]{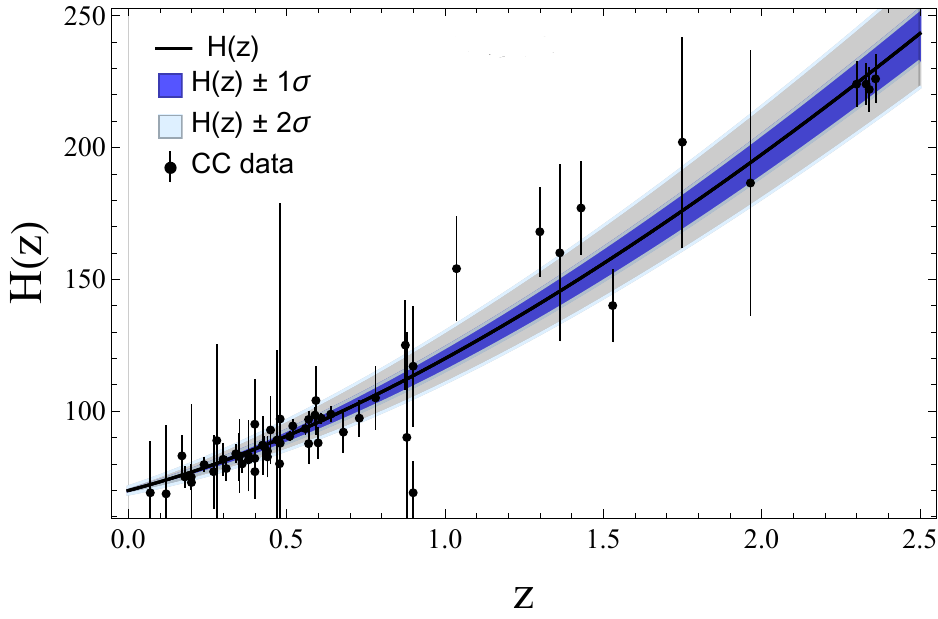}
    
    \end{minipage}
    \hspace{0.05\textwidth}
    \begin{minipage}{0.45\textwidth}
        \centering
        \includegraphics[width=\textwidth]{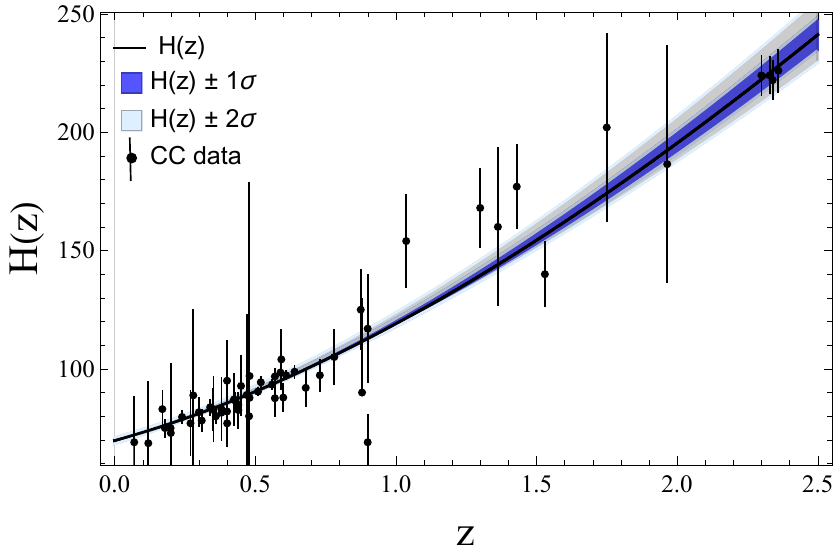}
    
    \end{minipage}
    \caption{The figures represent the predicted $H(z)$ function with 1$\sigma$ and 2$\sigma$ errors for Power-law (left) and Exponential (right) models.}
    \label{fig:figures2}
\end{figure}

\begin{figure}[h!]
    \centering
    \begin{minipage}{0.45\textwidth}
        \centering
        \includegraphics[width=\textwidth]{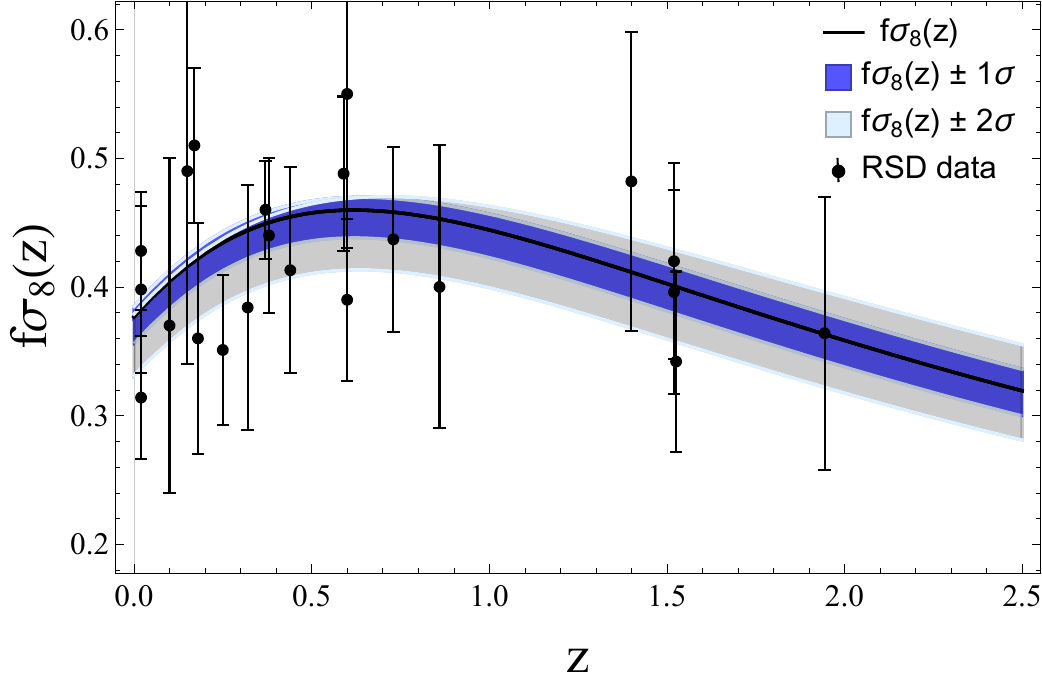}
       
    \end{minipage}
    \hspace{0.05\textwidth}
    \begin{minipage}{0.45\textwidth}
        \centering
        \includegraphics[width=\textwidth]{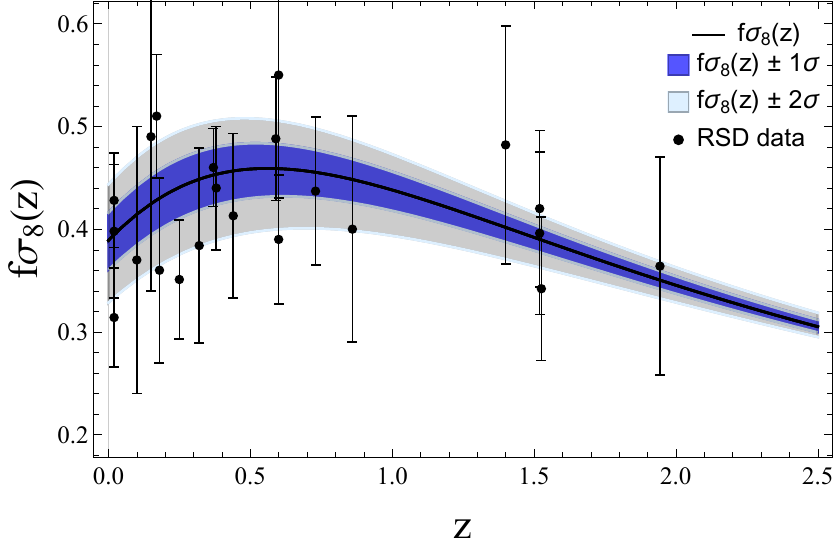}
    
    \end{minipage}
    \caption{The figures represent the predicted f$\sigma_{8}(z)$ function with 1$\sigma$ and 2$\sigma$ errors for Power-law (left) and Exponential (right) models.}
    \label{fig:figures3}
\end{figure}

\begin{figure}[h!]
    \centering
    \begin{minipage}{0.45\textwidth}
        \centering
        \includegraphics[width=\textwidth]{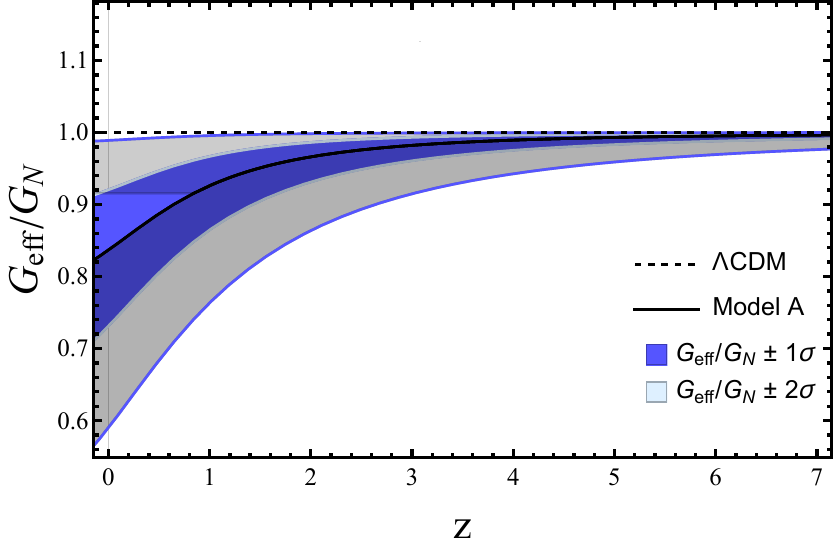}
    \end{minipage}
    \hspace{0.05\textwidth}
    \begin{minipage}{0.45\textwidth}
        \centering
        \includegraphics[width=\textwidth]{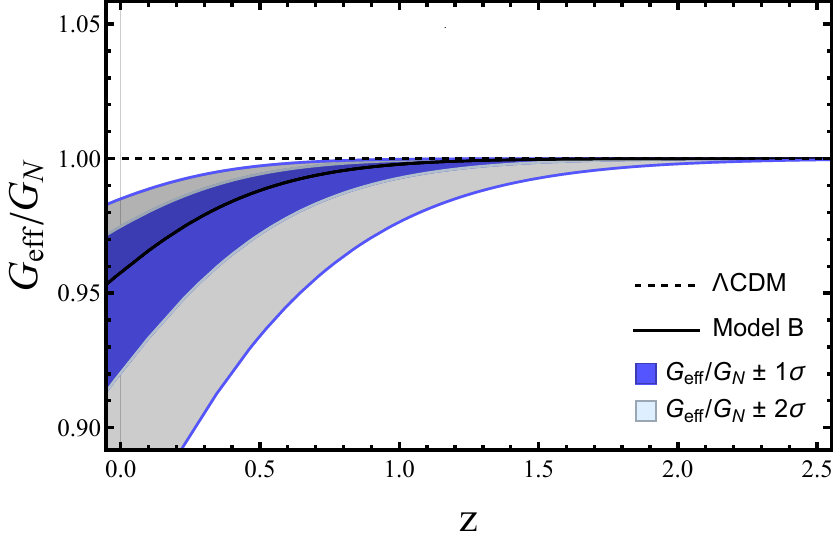}
        \label{fig:GB}
    \end{minipage}
        \caption{The evolution of the ratio $G_{eff}/G_{N}$ versus $z$, with their 1$\sigma$ and  2$\sigma$, for Power-law (left) and Exponential (right) models.}
    \label{G7}
\end{figure}

The 1D and 2D posterior distributions, representing the 68.3\% (1\(\sigma\)) and 95.4\% (2\(\sigma\)) confidence levels, are depicted in Figs. \ref{cc2} and \ref{cc3} for the Power-law and Exponential \( f(Q) \) models, respectively. For the Power-law model, Fig. \ref{cc2} shows a positive correlation in (\(\lambda\), \(\sigma_8\)) plan and negative correlations in (\(\sigma_8\), \(\Omega_m\)), (\(\Omega_m\), \(h\)), and (\(\Omega_m\), \(\lambda\)) plans. In the case of the Exponential model, Fig. \ref{cc2} illustrates a positive correlation in (\(\Omega_m\), \(\beta\)) plan, while (\(\Omega_m\), \(h\)) and (\(\Omega_m\), \(\sigma_8\)) plans exhibit negative correlations. Table \ref{CCT} represent the 1D and 2D posterior distributions for three models, $\Lambda$CDM (in green), Power-law (in blue), and Exponential (in red).\\

As an important evaluation of the results presented previously from MCMC analysis, the obtained values for the constrained parameters $\{h, \Omega_m, \sigma_8, \lambda, \beta\}$
 with Pantheon$^+$ +H(z)+RSD data, are used to draw the evolution of the distance modulus SNIa, Hubble Expansion rate, and the growth rate, with their 1$\sigma$ and 2$\sigma$ uncertainties, in terms of the redshift, $z$, as predicted by our $f(Q)$ models. This is shown in Figs. \ref{fig:figures1}, \ref{fig:figures2}, and \ref{fig:figures3}, where the $f(Q)$ predictions (black line) are compared with each sample data (black dots with the vertical line indicating the corresponding data uncertainty). It is evident that both models accurately reproduce the observations. Additionally, we plot the evolution of $G_{\text{eff}}/G_{\text{N}}$ for both models, including their 1$\sigma$ and 2$\sigma$ uncertainties, in Fig. \ref{G7}. For $f(Q)$ gravity, the expression for $G_{\text{eff}}/G_{\text{N}}$ is given by \cite{Geff}

\begin{equation}
    \frac{G_{\text{eff}}}{G_{\text{N}}} = \frac{1}{1+F_{Q}}
\end{equation}

In GR, this quantity is equal to 1. It allows us to quantify the deviation of $f(Q)$ models from GR. We find that $G_{\text{eff}}/G_{\text{N}}$ takes values of 0.82 and 0.96 for the Power-law and Exponential models of $f(Q)$, respectively. This indicates that the effective gravitational constant in the Power-law model is 18\% weaker than in the $\Lambda$CDM model, while in the Exponential model, it is only 4\% slightly weaker than $\Lambda$CDM. However, both models approach 1 in the past.

\section{Conclusions}\label{conclusion}
In this paper, we have performed a statistical analysis of a viable modify gravity, known as $f(Q)$ gravity model, which includes power-law, $f(Q)=Q+\alpha(Q/Q_0)^\lambda$, and Exponential, $f(Q)=Q+\alpha Q_0(1-\exp(-\beta\sqrt{Q/Q_0}))$, functions of the non-metricity $Q$.  The Power-law  function is chosen so that we  recover the $\Lambda$CDM expansion history of the Universe for $\lambda=0$ while the  Exponential function is so that it reduces to the symmetric teleparallel theory equivalent to GR without a cosmological constant for $\beta=0$.  
To observationally constrain these models, we used the Pantheon$^+$ compilation and 57 data points from Hubble measurements, along with redshift space distortion data. After numerically solving the modified Friedmann equations for the two models (\ref{Model1}) and (\ref{Model2}), we extracted the mean values of the model parameters using Markov Chain Monte Carlo. For the Power-law (Exponential) model, we found $\Omega_m = 0.256 \pm 0.014 \, (0.259 \pm 0.013)$, $h = 0.698 \pm 0.0068 \, (0.697 \pm 0.0069)$ km/s/Mpc, and $\sigma_8 = 0.866^{+0.039}_{-0.051} \, (0.834 \pm 0.029)$. Additionally, we found $\lambda = 0.172 \pm 0.079$ for the Power-law model and $\beta = 3.53^{+0.62}_{-0.77}$ for the Exponential model.\\

A comparative study between the \( f(Q) \) gravity and the \(\Lambda\)CDM models has been performed using the corrected Akaike Information Criterion and the Bayesian Information Criterion. The results of this comparison are shown in Table \ref{tab:model_comparison}. This table clearly shows that, according to the AIC\(_c\) criterion, the preferred models are, in order: the Exponential model, the Power-law model, and then the \(\Lambda\)CDM model. In other words, it turns out that the Exponential model fits the Pantheon$^+$ +H(z)+RSD datasets better than the \(\Lambda\)CDM model, indicating that \( f(Q) \) gravity has greater potential to explain the present-day dynamics of the universe than General Relativity. However, according to the BIC criterion, the preferred models are \(\Lambda\)CDM, the Exponential model, and then the Power-law model. Both AIC\(_c\) and BIC criteria prefer the Exponential model over the Power-law model.\\

Using the mean value parameters obtained previously in Table \ref{bf}, we have analyzed the ratio \( G_{\text{eff}} / G_N \), which quantifies the deviation of \( f(Q) \) models from General Relativity. Our analysis reveals that the effective gravitational constant in the Power-law model is 18\% weaker compared to the \(\Lambda\)CDM model. In contrast, the Exponential model shows a deviation where the effective gravitational constant is only 4\% slightly weaker than that of the \(\Lambda\)CDM model. Thus, It is crucial to investigate the constraints imposed by local gravity at the solar system level on the theory and determine their impact on the free parameters. Assessing the compatibility of these constraints with cosmological observations is also important. Additionally, examining the Newtonian and post-Newtonian limits can provide valuable physical constraints from various astrophysical observations.

\end{document}